\documentclass{amsart}
\usepackage{amsmath,amsfonts,amsthm,amssymb}

\usepackage{graphicx}
\theoremstyle{definition}

\theoremstyle{remark}

\numberwithin{equation}{section}

\begin{document}

\title[Dispersive dissipative string]{Hamiltonian extension and eigenfunctions for a time dispersive dissipative string}%
\author{Alexander Figotin}
\address{Department of Mathematics, University of California at Irvine, Irvine, CA 92697}

\author{Jeffrey Schenker}%
\address{Department of Mathematics, Michigan State University, East Lansing, MI 48824}
\curraddr{School of Mathematics, Institute for Advanced Study, Princeton, NJ 08540}%
\email{}%


\dedicatory{Dedicated to S. Molchanov on the occasion of his sixty-fifth birthday.}%
\begin{abstract}
We carry out a detailed analysis of a time dispersive 
dissipative (TDD) string, using our recently developed 
conservative and Hamiltonian extensions of TDD systems.  This
analysis of the TDD string includes, in particular: (i) an explicit
construction of its conservative Hamiltonian extension, consisting
of the physical string coupled to  ``hidden strings;" (ii) explicit
formulas for energy and momentum densities in the extended system,
providing a transparent physical picture accounting precisely for
the dispersion and dissipation; (iii) the eigenmodes for the
extended string system, which provide an eigenmode expansion for
solutions to the TDD wave equation for the TDD string. In particular, we find that in an eigenmode for the extended system the displacement of the physical string does not satisfy the formal eigenvalue problem, but rather an equation with a source term resulting from the excitation of the hidden strings.  The obtained results provide a solid basis for the rigorous treatment of the long standing problem of scattering by a TDD scatterer, illustrated here by the computation of scattering states for a string with dissipation restricted to a half line.
\end{abstract}
\maketitle
\section{Introduction}
The need for a Hamiltonian description of dissipative systems has
long been known, having been emphasized by Morse and Feshbach
\cite[Ch 3.2]{MorseFeshbach} forty years ago. Recently we have
introduced conservative \cite{FS1} and then Hamiltonian
\cite{FS2,FS3} theories of time dispersive and dissipative (TDD)
systems addressing that need. The extended Hamiltonian is
constructed by coupling the given TDD system to a system of ``hidden
strings" in a canonical way so that it has a transparent
interpretation as the system energy. It turns out that such a system
of hidden strings is, in fact, a canonical heat bath as described in
\cite[Section 2]{JP2}, \cite[ Section 2]{RB}.

This work is intended as a self contained companion to the papers
\cite{FS1, FS2, FS3}, illustrating the phenomena described there
through explicit calculations for the typical example of a TDD
string.

To keep everything elementary, let us consider a scalar wave
equation in one spatial dimension, such as might be used to describe
wave propagation along a homogeneous string.  In the absence of
dispersion, the displacement $\phi(x,t)$ of the string  at position
$x \in \mathbb R$ and time $t \in \mathbb R$ evolves by the $1D$
wave equation
\begin{equation}\label{eq:waveeqn}
\partial_t^2 \phi(x,t) \ - \ \gamma \partial_x^2 \phi(x,t) \ = \ f(x,t), \quad x, t \in \mathbb R,
\end{equation}
where
\begin{itemize}
\item $\gamma$ is the
tension of the string.
\item we have taken units in which the mass
per unit length is $1$.
\item $f(x,t)$ is an external \emph{driving force} per unit length, usually supposed bounded
and compactly supported.
\end{itemize}
For the most part, we consider \eqref{eq:waveeqn} with the string
\emph{at rest} at $t=-\infty$,
\begin{equation}
\lim_{t \rightarrow -\infty} \phi(x,t) \ = \ \lim_{t \rightarrow
\infty} \partial_t \phi(x,t) \ = \ 0.
\end{equation}
Thus the solution $\phi(x,t)$ is a function of the driving force
$f(x,t)$, indeed
 \begin{equation}\label{eq:drivenstring}
 \phi(x,t) \ = \ \frac{1}{2 v } \int_{-\infty}^t  \int_{x- v (t-t')}^{x + v (t -t')}
 f(y, t') \, \mathrm d y \, \mathrm d t'
 \end{equation}
 with $v = 1/ \sqrt{\gamma}$ the speed of propagation on the string.

 In this paper we consider a
modification of \eqref{eq:waveeqn} incorporating friction in the
form of a dissipative term with time dispersion. Specifically, we
consider the following equation
\begin{equation}\label{eq:TDDwaveeqn}
\partial_t \left \{ \partial_t \phi(x,t) + \int_{-\infty}^t \chi(x,t-t') \partial_{t'} \phi(x,t') \, \mathrm d t'
\right \} - \gamma \partial_x^2 \phi(x,t) \ = \ f(x,t) ,
\end{equation}
with $\chi(x,\tau)$ a given function, called the susceptibility,
satisfying a power dissipation condition \textemdash \
\eqref{eq:PDC} below.

The physical idea behind \eqref{eq:TDDwaveeqn} is as follows.  The
wave equation \eqref{eq:waveeqn} can be expressed in terms of the
string momentum $\pi(x,t) = \partial_t \phi(x,t)$ as
\begin{equation}\label{eq:F=dp}
\partial_t \pi(x,t) \ = \ \text{sum of all forces at $x$} \ = \ \underbrace{f(x,t)}_{\text{driving force}} \ + \ \underbrace{\gamma \partial_x^2 \phi(x,t)}_{\text{string tensile force}}.
\end{equation}
In the damped string \eqref{eq:TDDwaveeqn}, the basic relation
$\partial_t \pi = $ ``sum of all forces'' still holds, but the
simple relationship $\pi = \partial_t \phi$ between the string
momentum and velocity is replaced by the \emph{material relation}
\begin{equation}\label{eq:TDDp=v}
\pi(x,t) \ = \  \partial_t \phi(x,t) + \int_{-\infty}^t \chi(x,t-t')
\partial_{t'} \phi(x,t') \, \mathrm d t' .
\end{equation}
Eq.\ \eqref{eq:TDDp=v} is supposed to be a phenomenological relation
describing the interaction of the string with a surrounding medium,
expressing the fact that some of the string momentum is absorbed by
the medium and then partially retransmitted to the string with
delay.

\begin{figure} [ptb]
\begin{center}
\includegraphics[ height=2.15in, width=4in ]{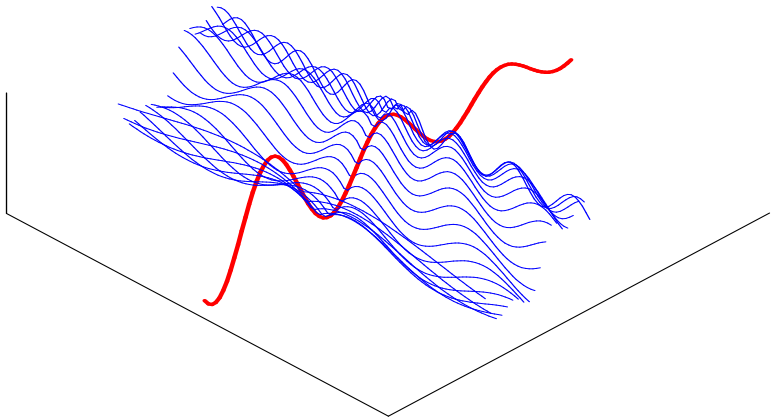} \caption{The extended system
consists of the physical string (thick/red) coupled to an
independent ``hidden string'' (thin/blue) at every point of
dissipation. This Hamiltonian system models exactly the dissipation
and dispersion in the physical string via the exchange of energy
with the hidden strings. } \label{fig:strings}
\end{center}
\end{figure}

The wave equation \eqref{eq:waveeqn} may be expressed as a
Hamiltonian system. As described in \cite{FS2,FS3}, the procedure of
going from \eqref{eq:waveeqn} to \eqref{eq:TDDwaveeqn} is naturally
understood in this context and indeed \eqref{eq:TDDwaveeqn}
\emph{can be derived from a larger Hamiltonian system, with
additional variables .} The additional variables of this extension
may be interpreted as describing a ``hidden string,'' with internal
coordinate $s$, attached to each point $x$ of the physical string
(see figure \ref{fig:strings}).
These ``hidden strings'' are coupled to the physical string via a
coupling function $\varsigma(x,s)$, but do not interact directly
with one another.\footnote{The lack of interaction between hidden
strings is not essential.  Indeed the methods of \cite{FS1,FS2,FS3}
allow to consider systems with ``spatial dispersion'' in the
material relation
$$ \pi(x,t) \ = \ \partial_t \phi(x,t) + \int_{\mathbb R} \int_{0}^\infty \chi(x',\tau)
\partial_t \phi(x-x',t-\tau)\, \mathrm d \tau \, \mathrm d x'.
$$
The extensions for such systems involve interaction between the
hidden strings, but for simplicity we do not pursue this here.} Upon
``integrating out'' the hidden strings, the result is an effective
dynamics of the form \eqref{eq:TDDwaveeqn} for the physical string,
with susceptibility
\begin{equation}\label{eq:susceptibility=sigma2}
\chi(x,\tau) \ = \ \frac{1}{2} \int_{\mathbb R} \varsigma(s,x)
\int_{s-\tau}^{s + \tau} \varsigma(x,r) \, \mathrm d r \, \mathrm d
s .
\end{equation}
Different choices of the coupling function $\varsigma(x,s)$ produce
different susceptibilities $\chi(x,\tau)$, and furthermore
\cite{FS2,FS3} any susceptibility satisfying the power dissipation
condition \textemdash \ \eqref{eq:PDC} below \textemdash \ can be
obtained in this way.

Given the role of the TDD wave equation \eqref{eq:TDDwaveeqn} as a
phenomenological description for the interaction of the string with
its environment, it is not surprising that an extended system exists
which produces the effective dynamics \eqref{eq:TDDwaveeqn}, at
least approximately. However, the main point of \cite{FS2,FS3},
which we explain below, is that it is possible to construct
\emph{explicitly} an extended system which \emph{exactly} reproduces
\eqref{eq:TDDwaveeqn}. Further, using that system we can derive
simple expressions for quantities like the energy and wave momentum
densities for the damped string in terms of the susceptibility
$\chi$.

In addition, as illustrated below, the extended system clarifies the
nature and role of eigenfunctions in a TDD system by providing a
natural eigenmode expansion for solutions to \eqref{eq:TDDwaveeqn}.
In this context, an eigenfunction $\exp(-\mathrm i \omega t)
\phi_\omega(x)$ is a time periodic steady state solution to
\eqref{eq:TDDwaveeqn}. For the simple non-dissipative string
\eqref{eq:waveeqn}, an eigenmode solves
\begin{equation}
  \gamma \partial_x^2 \phi_\omega(x) \ = \ - \omega^2 \phi_\omega(x) ,
\end{equation}
which is to say it is a linear superposition of the plane waves
$\exp(\pm \mathrm i \omega x/ \sqrt{\gamma})$. For the TDD string
\eqref{eq:TDDwaveeqn}, the dissipation induced by the dispersion in
\eqref{eq:TDDp=v} may preclude the existence of a steady state
solution without a source term. However, there is no difficulty in
constructing the eigenmodes for the \emph{extended system}, as it is
without dissipation.  We find below that the resulting eigenmode
equation for $\phi_\omega$
\begin{equation}\label{eq:introeigenfunctionequation}
  \gamma \partial_x^2\phi_\omega(x) \ = \ - \omega^2 \left (1 + \mathrm{Re} \, \widehat
  \chi(x,\omega) \right ) \phi_\omega(x)  - \mathrm i \omega^2 \mathrm{Im} \,
  \widehat \chi(x,\omega) g_\omega(x) ,
\end{equation}
has a source term involving an \emph{arbitrary} function $g_\omega$,
which depends on the configuration of the hidden strings. Here
$\widehat \chi$ is the $\tau$-Fourier transform of $\chi$,
\begin{equation}\label{eq:FTchi}
  \widehat \chi(x,\omega) \ = \ \int_0^\infty \mathrm e^{\mathrm i \omega
  \tau} \chi(x,\tau) \, \mathrm d \tau.
\end{equation}

Let us now sketch the construction of the Hamiltonian extension. To
start we write \eqref{eq:waveeqn} as a first order system,
introducing the momentum density $\pi(x,t) \ =\partial_t \phi(x,t)$,
so that
\begin{equation}
\begin{aligned}\label{eq:firstorder}
  \partial_t \phi(x,t) \ &= \  \pi(x,t)  \\
  \partial_t \pi(x,t) \ &= \ \gamma \partial_x^2 \phi(x,t)  + f(x,t).
\end{aligned}
\end{equation}
This first order system is Hamiltonian, with symplectic form
\begin{equation}
J ( \phi_1, \pi_1; \phi_2, \pi_2) \ = \ \int_{\mathbb R} (\phi_1(x)
\pi_2(x) - \phi_2(x) \pi_1(x) ) \, \mathrm d x
\end{equation}
and time dependent Hamilton function
\begin{equation}
H_f(\phi,\pi,t) \ = \ \frac{1}{2} \int_{\mathbb R} \left ( \pi(x)^2
+ \gamma |\partial_x \phi(x)|^2 \right ) \, \mathrm d x  -
\int_{\mathbb R} f(x,t) \phi(x) \, \mathrm d x.
\end{equation}
As the system is non-autonomous, $H_f(\pi(\cdot,t),\phi(\cdot,t),t)$
is not conserved.  Instead along a trajectory to
\eqref{eq:firstorder}, we have
\begin{equation}\label{eq:nonautdHdt}
  \frac{\mathrm d H_f}{\mathrm d t} \ = \ \partial_t H_f(\phi(\cdot,t),\pi(\cdot,t),t)
  \ = \ - \int_{\mathbb R} \partial_t f(x,t) \phi(x,t) \, \mathrm d x .
\end{equation}
We take the internal energy of the (undamped) string to be the
Hamilton function $H_0(\phi,\pi)$ of the system with $f \equiv 0$.
Thus, using \eqref{eq:nonautdHdt} we obtain
\begin{equation}\label{eq:autdHdt}
  \frac{\mathrm d H_0}{\mathrm d t} \ = \ \frac{\mathrm d H_f}{\mathrm d
  t}+ \frac{\mathrm d  \ }{\mathrm d t}
  \int_{\mathbb R} f(x,t) \phi(x,t) \, \mathrm d x \ = \
  \int_{\mathbb R} f(x,t) \partial_t \phi(x,t) \, \mathrm d x .
\end{equation}
In other words, the rate of work done on the system is the integral
of the external force times the velocity -- ``power $=$ force
$\times$ velocity.''

Observe that the Hamiltonian $H_0$ can be written
\begin{equation}\label{eq:HaminK}
H_0 (\phi,\pi) \ = \ \frac{1}{2} \left  ( \| K_\pi \pi \|
_{L^2(\mathbb R)}^2 + \| K_\phi \phi \| _{L^2(\mathbb R)}^2 \right )
\end{equation}
where
\begin{equation}
K_\phi \ = \ \sqrt{\gamma} \partial_x, \quad K_\pi \ = \
\text{Identity} .
\end{equation}
Based on this expression, we separate the equations of motion
\eqref{eq:firstorder}, into ``material relations''
\begin{align}\label{eq:fphi}
f_\phi(x) \ &= \ K_\phi \phi(x) \ = \ \sqrt{\gamma} \partial_x \phi(x) \\
f_\pi(x) \ &= \ K_\pi \pi(x) \ = \   \pi(x) , \label{eq:fpi}
\end{align}
and dynamical equations,
\begin{equation}\label{eq:firstorder2}
\begin{aligned}
\partial_t \phi(x,t) \ &= \ K_\pi^\dagger f_\pi(x,t)  \\
  \partial_t \pi(x,t) \ &= \ - K_\phi^\dagger f_\phi(x,t) + f(x,t) .
\end{aligned}
\end{equation}

To obtain a first order TDD system leading to \eqref{eq:TDDwaveeqn},
we follow \cite{FS2,FS3} and replace the identity (\ref{eq:fpi})
with the ``material relation,''
\begin{equation}
   f_\pi(x,t) + \int_0^\infty \chi(x,\tau) f_\pi(x,t-\tau) \, \mathrm d \tau \ = \
  \pi(x,t)\label{eq:fpiTDD}  ,
\end{equation}
but maintain \eqref{eq:fphi} and the equations of motion
\eqref{eq:firstorder2}, i.e.,
\begin{equation}\label{eq:TDDmotion}
\begin{aligned}
\partial_t \phi(x,t) \ &= \ f_\pi(x,t) \\
  \partial_t \pi(x,t) \ &= \  \gamma \partial_x^2 \phi(x,t)
  + f(x,t).
\end{aligned}
\end{equation}
(One could also introduce dispersion in the relation between $\phi$
and $f_\phi$, or even dispersion mixing $f_\phi$ and $f_\pi$, but
for simplicity we consider here only TDD in the momentum.) The key
requirements on  the \emph{susceptibility} $\chi$ are
\begin{enumerate}
\item Causality, manifest in (\ref{eq:fpiTDD}) in that the integral on the l.h.s.\ depends only on the past.
\item Power dissipation, which is the requirement that for every $x$
\begin{equation}\label{eq:PDC}
\int_{-\infty}^\infty g(t) \int_{0}^\infty \chi(x,\tau) \partial_t
g(t - \tau) \,\mathrm{d} \tau \, \mathrm{d} t
 \ \ge \ 0
\end{equation}
for an arbitrary function $g(t)$, say compactly supported.
\end{enumerate}

The significance of the power dissipation condition is as follows.
We suppose that the internal energy of the TDD system at time $t$ is
given by
\begin{equation}\label{eq:Haminf}
  H_0 \ = \ \frac{1}{2} \int_{\mathbb R} \left ( f_\pi(x,t)^2 + \gamma
  \partial_x \phi(x,t)^2 \right ) \, \mathrm d x ,
\end{equation}
with $f_\pi(x,t)$ as in \eqref{eq:fpiTDD}.  Then
\begin{equation}\label{eq:diftH}
\begin{aligned}
&\frac{\mathrm d}{\mathrm d t} H_0(t) \\ &= \ \int_{\mathbb R} \Bigl
( \gamma  \partial_x \phi (x,t) \partial_t \partial_x \phi (x,t)
 \ + \ f_\pi(x,t) \partial_t f_\pi (x,t) \Bigr ) \, \mathrm d x \\
&= \ \int_{\mathbb R} \Bigl ( \gamma \partial_x \phi (x,t)
\partial_x f_\pi (x,t) \ + \
f_\pi(x,t)  \partial_t \pi (x,t)
\\ & \quad  \quad \quad -  f_\pi (x,t)  \int_0^\infty\chi(x,\tau)  \partial_t
f_\pi(x,t-\tau) \,
\mathrm{d} \tau \Bigr )  \, \mathrm d x\\
&= \ \int_{\mathbb R} \Bigl ( \gamma \partial_x \phi (x,t)
\partial_x f_\pi (x,t) \ + \ \gamma
f_\pi(x,t)  \partial_x^2 \phi (x,t)
\\ & \quad  \quad \quad \ + \  f_\pi(x,t) f(x,t) \ - \  f_\pi (x,t)
\int_0^\infty\chi(x,\tau)  \partial_t f_\pi(x,t-\tau) \,
\mathrm{d} \tau \Bigr )  \, \mathrm d x \\
 &= \  \int_{\mathbb R}  f(x,t) \partial_t \phi(x,t) \, \mathrm d x \ - \
 \int_{\mathbb R}   \left [ \int_0^\infty\chi(x,\tau)
 \partial_t^2 \phi(x,t-\tau) \,
\mathrm{d} \tau  \right ]  \partial_t \phi(x,t) \, \mathrm d x ,
\end{aligned}
\end{equation}
where in going from the third to the final line we have used
integration by parts and the assumption that $\partial_x\phi$ and
$f_\pi$ vanish at spatial infinity. The first term on the r.h.s.\ is
the rate of work done on the system by the external force $f$.
Similarly, we interpret the second term as the power loss of the
dissipative force, $-\int \chi(\tau)
\partial_t^2 \phi(x,t -\tau) \mathrm d \tau$.  Thus, power dissipation
amounts to the requirement that, for any trajectory, the \emph{total
work done by the dissipative force is negative}.

Let us note that the right hand side of \eqref{eq:PDC} is equal to
\begin{multline}\label{eq:rhsPDC} \chi(x,0) \int_{-\infty}^\infty g(t)^2 \, \mathrm d
t + \int_{-\infty}^\infty g(t) \int_{0}^\infty \partial_\tau
\chi(x,\tau) g(t - \tau) \,\mathrm{d} \tau \, \mathrm{d} t
 \\ = \
\int_{\mathbb R^2} \Bigl \{  \chi(x,0) \delta(t_1 - t_2) +
\frac{1}{2}[\partial_{\tau} \chi](x,|t_1-t_2|) \Bigr \} g(t_1)
g(t_2) \,  \mathrm d^2 t ,
\end{multline}
which follows from integration by parts after noting that
$\partial_t g(t-\tau) = - \partial_\tau g(t-\tau)$. Here $\delta(t)$
is the Dirac delta function. Thus the power dissipation condition is
equivalent to the statement that for every $x$ the (generalized)
function
\begin{equation}\label{eq:FrictFunc}
D_{\chi}(x,t) \ = \  \chi(x,0) \delta(t) +
\frac{1}{2}[\partial_{\tau} \chi](x,|t|)
\end{equation}
is \emph{positive definite} in the sense of the classical Bochner's
Theorem, see \cite[Theorem IX.9]{RS1}. Thus, by Bochner's Theorem,
the time Fourier transform of $D_\chi$ is a non-negative measure,
\begin{multline}\label{eq:FrictFuncFT}
\widehat D_\chi (x,\omega) \ = \  \chi(x,0) + \frac{1}{2}
\int_{-\infty}^\infty [\partial_{\tau} \chi](x,|t|) \mathrm
e^{\mathrm i \omega t} \, \mathrm d t \\ = \  \chi(x,0) +
\int_{0}^\infty \partial_{\tau} \chi(x,\tau) \cos(\omega \tau) \,
\mathrm d \tau \\
 = \  \omega \int_0^\infty \chi(x,\tau) \sin(\omega
\tau) \, \mathrm d \tau \ = \ \omega \mathrm{Im} \widehat
\chi(x,\omega) \ \ge \ 0 .
\end{multline}
Since $\chi$ is real, $\widehat
D_{\chi}(x,\omega)$ is symmetric under $\omega \leftrightarrow
-\omega$, and indeed Bochner's Theorem shows that a symmetric
measure of suitably bounded growth is non-negative if and only if it
is the Fourier transform of a real positive definite distribution.

The simplest physically relevant example of a susceptibility
satisfying the power dissipation condition is obtained with
$\chi(x,\tau) = \alpha > 0$, a positive constant. Then
\begin{equation}
  f_\pi(x,t) + \alpha \int_0^\infty f_\pi(x,t - \tau) \, \mathrm{d} \tau \ = \
  \pi(x,t) .
\end{equation}
Using the equation of motion $f_\pi = \partial_t \phi$, we find that
\begin{equation}
\begin{aligned}
  \pi(x,t) \ &= \
 \partial_t \phi(x,t)  + \alpha \int_0^\infty \partial_t
  \phi(x,t-\tau)\, \mathrm d \tau  \\
  \ &= \  \partial_t \phi(x,t) + \alpha  \phi(x,t) ,
\end{aligned}
\end{equation}
where we have applied the boundary condition $\lim_{t \rightarrow
-\infty} \phi(x,t) = 0$. Combined with the equation of motion
$\partial_t \pi = \gamma \partial_x^2 \phi + f$ we obtain
\begin{equation}
 \partial_t^2 \phi(x,t) + \alpha\partial_t \phi(x,t) -
\gamma\partial_x^2 \phi(x,t) \ = \ f(x,t) ,
\end{equation}
which is the dynamical equation for a driven damped string, with
damping force per unit length $- \alpha
\partial_t \phi(x,t).$ Note that $\alpha$ is dimensionally an inverse time
and $1/\alpha$ is the characteristic time for the damping of
oscillations.

A more realistic model for damping is obtained by supposing
$\chi(x,\tau)$ to be a non-trivial function of $\tau$ as in
(\ref{eq:fpiTDD}). To allow for damping restricted to only a part of
the string, we suppose that $\chi$ depends on $x$ as well. For
instance, we could take the Debye susceptibility
\begin{equation}
\chi(x,\tau) \ = \alpha(x)\mathrm e^{-\nu(x) \tau} ,
\end{equation}
with $\alpha(x) \ge 0$ and $\nu(x) \ge 0$ non-negative functions of
$x$. This results in the following integral-differential equation
for $\phi$
\begin{equation}
\partial_t^2 \phi(x,t) + \alpha(x)  \int_0^\infty \mathrm e^{-\nu(x) \tau} \partial_t^2
  \phi(x,t-\tau) \, \mathrm d \tau -
\gamma\partial_x^2 \phi(x,t) \ = \ f(x,t) ,
\end{equation}
or after integration by parts
\begin{multline}
  \partial_t^2 \phi(x,t) + \alpha(x)
  \partial_t \phi(x,t) -
  \alpha(x)\nu(x) \phi(x,t) \\ + \alpha(x) \nu(x)^2 \int_0^\infty \mathrm e^{-\nu(x) \tau}
  \phi(x,t-\tau) \, \mathrm d \tau -
\gamma\partial_x^2 \phi(x,t) \ = \ f(x,t) .
\end{multline}
Then
\begin{equation}\label{eq:DebyeFrictFunc}
\widehat D_{\chi}(x,\omega) \ = \ \alpha(x) \int_0^\infty \mathrm
e^{-\nu(x) \tau} \omega \sin (\omega \tau) \, \mathrm d \tau \ = \
\alpha(x) \frac{\omega^2}{\nu(x)^2 + \omega^2} \ \ge \ 0 ,
\end{equation}
so the Debye susceptibility satisfies the power dissipation
condition.

The results of \cite{FS2,FS3} show that under the above conditions
it is possible to find a coupling function $\varsigma(x,s)$, $s \in
\mathbb R$, such that solutions to the TDD equations are generated
by solutions to the following extended system
\begin{equation}\label{eq:extended}
\begin{aligned}
\partial_t \phi(x,t)  \ &= \ f_\pi(x,t)  \\
\partial_t \pi(x,t) \ &= \ \gamma \partial_x^2 \phi(x,t) + f(x,t)\\
\partial_t \psi(x,s,t) \ &= \ \theta(x,s,t) \\
\partial_t \theta(x,s,t) \ &= \ \partial_s^2 \psi(x,s,t) + \varsigma(x,s) f_\pi(x,t) ,
\end{aligned}
\end{equation}
with
\begin{equation}\label{eq:extendedfpi}
f_\pi(x,t) \ = \ \pi(x,t) - \int_{-\infty}^\infty \varsigma(x,s)
\psi(x,s,t) \,  \mathrm d s.
\end{equation}
That is, given a solution $(\phi,\pi,\psi,\theta)$ to
\eqref{eq:extended}, at rest at $t=-\infty$ and with $f_\pi$ given
by \eqref{eq:extendedfpi}, the first two coordinates $(\phi,\pi)$
obey \eqref{eq:TDDmotion} with $f_\pi$ given by \eqref{eq:fpiTDD},
and conversely any solution to (\ref{eq:fpiTDD}, \ref{eq:TDDmotion})
with the string at rest at $t=-\infty$ arises in this way.  Note
that for each fixed $x$ the additional variables $\theta, \psi$ may
be interpreted as describing the oscillations of a ``hidden string''
with coordinate $s$, displacement $\psi(x,s,t)$, momentum
$\theta(x,s,t)$ and driven by external force $\varsigma(x,s)
f_\pi(x,t)$. Thus \eqref{eq:extended} is precisely the system of
coupled strings described above and illustrated in Figure
\ref{fig:strings}.

The extended system, consisting of the physical and hidden strings
is Hamiltonian with symplectic form
\begin{multline}
\mathcal
J(\phi_1,\pi_1,\psi_1,\theta_1;\phi_2,\pi_2,\psi_2,\theta_2) \
 = \ \int_{\mathbb R} \left \{ \phi_1(x)
\pi_2(x) - \phi_2(x) \pi_1(x) \right \}  \mathrm d x \\ +
\int_{\mathbb R^2} \left \{ \psi_1(x,s) \theta_2(x,s) - \psi_2(x,s)
\theta_1(x,s) \right \}  \mathrm d s \, \mathrm d x
\end{multline}
and Hamilton function
\begin{multline}\label{eq:extendedHam}
  \mathcal H_f(\phi,\pi,\psi,\theta,t) \ = \ \frac{1}{2}\int_{\mathbb R} \left \{ f_\pi(x)^2 +
  \gamma ( \partial_x \phi(x) )^2 \right \} \mathrm d
  x - \int_{\mathbb R} f(x,t) \phi(x) \, \mathrm d x \\
  + \frac{1}{2} \int_{\mathbb R^2} \left \{ \theta(x,s)^2 +
  (\partial_s \psi(x,s))^2  \right \} \mathrm d s \, \mathrm d
  x ,
\end{multline}
where $f_\pi$ is given by \eqref{eq:extendedfpi}, that is $f_\pi(x)
\ = \ \pi(x) - \int \varsigma(x,s) \psi(x,s) \, \mathrm d s$.

The dynamical equation for an excitation of the hidden string at $x$
is a driven wave equation
\begin{equation}\label{eq:hiddenstringwaveeqn}
  \partial_t^2 \psi(x,s,t) \ = \ \partial_s^2 \psi(x,s,t) +
  \varsigma(x,s) f_\pi(x,t).
\end{equation}
The solution to \eqref{eq:hiddenstringwaveeqn} with the hidden
string at rest at $t=-\infty$ is easily written down, see
\eqref{eq:drivenstring}:
\begin{equation}
  \psi(x,s,t) \ = \ \frac{1}{2} \int_0^\infty \int_{s-\tau}^{s+\tau}
  \varsigma(x,r) \, \mathrm d r \, f_\pi(x,t-\tau)  \, \mathrm d \tau .
\end{equation}
Thus
\begin{multline}\label{eq:fundamental}
  \int_{\mathbb R} \varsigma(x,s) \psi(x,s,t) \, \mathrm d s
  \\ = \ \int_0^\infty \left [\frac{1}{2} \int_{\mathbb R} \varsigma(x,s)
  \int_{s-\tau}^{s+\tau}
  \varsigma(x,r) \, \mathrm d r \, \mathrm d s \right ]
  f_\pi(x,t-\tau) \,
  \mathrm d \tau .
\end{multline}
Comparing \eqref{eq:fpiTDD} and \eqref{eq:extendedfpi} we see that
for the extension to reproduce the TDD system upon elimination of
$\psi$ and $\theta$ it is necessary and sufficient that the coupling
$\varsigma(x,s)$ satisfy \eqref{eq:susceptibility=sigma2}, that is
\begin{equation}\label{eq:susceptibility=sigma2again}
\chi(x,\tau) \ = \ \frac{1}{2} \int_{\mathbb R} \varsigma(s,x)
\int_{s-\tau}^{s + \tau} \varsigma(x,r) \, \mathrm d r \, \mathrm d
s .
\end{equation}
The existence of such a function, which is unique under a natural
symmetry assumption, is guaranteed by the power dissipation
condition \cite{FS2,FS3}.

Indeed, if we let $\widehat \varsigma (x,\sigma)$ denote the
$s$-Fourier transform of $\varsigma$,
\begin{equation}
  \widehat \varsigma (x,\sigma) \ = \ \int_{\mathbb R}
  \varsigma(x,s) \mathrm e^{\mathrm i \sigma s} \, \mathrm d s,
\end{equation}
then \eqref{eq:susceptibility=sigma2again} is equivalent to
\begin{equation}\label{eq:fundamentalFT}
\widehat D_{\chi}(x,\omega) \ = \ \widehat \varsigma(x,\omega)
\widehat \varsigma(x,-\omega),
\end{equation}
so it suffices to take \cite{FS2}
\begin{equation}
  \widehat \varsigma (x,\sigma) \ = \ \sqrt{2 \widehat D_{\chi}(x,\sigma)}
  .
\end{equation}
Furthermore, this choice of $\widehat \varsigma(x,\sigma)$ is unique
if we ask further that $\widehat \varsigma(x,\sigma) \ge 0$ and that
$\sigma \mapsto \widehat \varsigma(x,\sigma)$ be symmetric. Then,
\begin{equation}\label{eq:formofsigma}
  \varsigma(x,s) \ = \ \frac{1}{2 \pi} \int_{\mathbb R}
  \sqrt{2 \widehat D_\chi(x,\sigma)} \mathrm e^{-\mathrm i \sigma s} \,
  \mathrm d \sigma \ = \ \frac{1}{2 \pi} \int_{\mathbb R}
  \sqrt{2 \widehat D_\chi(x,\sigma)} \cos(\sigma s) \, \mathrm d \sigma
  ,
\end{equation}
is real, symmetric, and positive definite.

For example, in view of \eqref{eq:FrictFuncFT},
\eqref{eq:DebyeFrictFunc}, and \eqref{eq:formofsigma}, the following
coupling produces the Debye susceptibility,
\begin{equation}
\varsigma(x,s) \ = \ \sqrt{2 \alpha(x)} \partial_s  \Psi(\nu(x) s) .
\end{equation}
where
\begin{multline}
\Psi(s) \ = \ \frac{1}{2 \pi \mathrm i} \int_{-\infty}^\infty
\frac{1}{\omega} \sqrt{\frac{\omega^2}{1 + \omega^2}}   \mathrm
e^{\mathrm i \omega s} \, \mathrm d \omega
\\ = \ \frac{s}{|s|}
\frac{1}{\pi} \int_0^1 \frac{1}{\sqrt{1-u^2}} \mathrm e^{-u|s|} \,
\mathrm d u \  = \ \frac{s}{|s|}  \frac{1}{\pi}
\int_0^{\frac{\pi}{2}} \mathrm e^{-|s| \sin(\phi) } \, \mathrm d
\phi .
\end{multline}
That is,
\begin{equation}
\varsigma(x,s) \ = \ \sqrt{2 \alpha(x)} \left ( \delta(s) -
\frac{\nu(x)}{\pi} \int_0^{\frac{\pi}{2}} \sin(\phi) \mathrm
e^{-\nu(x)|s| \sin(\phi) } \, \mathrm d \phi  \right ) .
\end{equation}

\section{Local energy and momentum conservation in the extended system}
We interpret the Hamiltonian $\mathcal H_0$ with $f \equiv 0$ as the
internal energy of the damped string system consisting of the
coupled physical and hidden strings. We have conservation of energy
in the extended system, in the form
\begin{equation}\label{eq:diftcalH}
  \frac{\mathrm d \ }{\mathrm d t} \mathcal H_0 \ = \ \int_{\mathbb
  R} f(x,t)\partial_t \phi(x,t) \, \mathrm d x ,
\end{equation}
i.e., the rate of change of $\mathcal H_0$ is the rate of work done
on the system by the external force.

A significant advantage of working with the extended system is a
transparent interpretation of the energy of the dissipative string
as a sum of contributions from the physical and hidden strings.
That is, it is natural to break the $\mathcal H_0$ into two pieces
\begin{equation}
  \mathcal H_0(\phi,\pi,\psi,\theta,t) \ = \ H_0(\phi, \pi,\psi)
  + H_{hs}(\psi,\theta) ,
\end{equation}
the energy of the physical string and hidden strings respectively,
\begin{align}
    H_0(\phi, \pi,\psi) \ &= \ \frac{1}{2} \int_{\mathbb R} \left \{ f_\pi(x)^2 +
    \gamma (\partial_x \phi(x))^2 \right \} \mathrm d x \\
    H_{hs}(\psi,\theta) \ &= \ \frac{1}{2} \int_{\mathbb R^2} \left \{ \theta(x,s)^2 +
  (\partial_s \psi(x,s))^2  \right \} \mathrm d s \mathrm d
  x ,
  \end{align}
with $f_\pi$ given by \eqref{eq:extendedfpi}.

The internal energy $\mathcal H_0$ can be written as the integral of
a local energy density
\begin{equation}
  \mathcal E (x,t) \ = \ E_0(x,t) + E_{hs}(x,t)
\end{equation}
with
\begin{equation}
  \begin{aligned}
    E_0(x,t) \ &= \ \frac{1}{2} \left \{ (\partial_t \phi(x,t))^2 + \gamma
    (\partial_x \phi(x,t))^2 \right \} \\
    E_{hs}(x,t) \ &= \ \frac{1}{2}  \int_{\mathbb R} \left \{
    (\partial_t \psi(x,s,t))^2 + (\partial_s \psi(x,s,t))^2 \right \} \mathrm
    d s .
  \end{aligned}
\end{equation}
The energy conservation law \eqref{eq:diftcalH} has the following
local expression
\begin{equation}\label{eq:localconservation}
  \partial_t \mathcal E(x,t)  + \partial_x J(x,t) \ = \ f(x,t)
  \partial_t \phi(x,t)
\end{equation}
with the energy current
\begin{equation}\label{eq:energyflux}
  J(x,t) \ = \ - \gamma f_\pi(x,t)
  \partial_x \phi(x,t) \ = \ - \gamma \partial_t \phi(x,t) \partial_x\phi(x,t).
\end{equation}

It is interesting to compute the time derivatives of $E_0$ and
$E_{hs}$ alone:
\begin{align}\label{eq:localE0}
\partial_t E_0(x,t) +  \partial_x  J(x,t) \ &= \ f(x,t) \partial_t \phi(x,t)  -
\partial_t \Delta(x,t) \partial_t \phi(x,t)  \\ \label{eq:localEhs}
\partial_t E_{hs}(x,t) \ &= \ \partial_t \Delta(x,t) \partial_t \phi(x,t)
\end{align}
with
\begin{multline}
\Delta(x,t) \ = \ \pi(x,t) - f_\pi(x,t) \\ = \ \int_{\mathbb R}
\varsigma(x,s) \psi(x,s,t)\, \mathrm d s \ = \ \int_0^\infty
\chi(x,\tau) \partial_t \phi(x,t-\tau) \, \mathrm d \tau ,
\end{multline}
where we have used \eqref{eq:fundamental}. The first of these
equations \eqref{eq:localE0} is simply the local version of the
energy law for the TDD string \eqref{eq:diftH}. From the second
\eqref{eq:localEhs}, we see that the energy of the hidden strings,
which is the energy lost to dissipation up to time $t$, is
\begin{equation}
\begin{aligned}
E_{hs}(x,t) \ &= \ \int_{-\infty}^t \partial_{t'} \phi(x,t')
\int_0^\infty \chi(x,\tau)
\partial_{t'}^2 \phi(x,t'-\tau) \, \mathrm d \tau \, \mathrm d
t' . \\
&= \ \int_{-\infty}^t \int_{-\infty}^t D_\chi(x,t_1 -t_2)
\partial_{t_1} \phi(x,t_1) \partial_{t_2} \phi(x,t_2) \, \mathrm d t_1 \, \mathrm d t_2
.
\end{aligned}
\end{equation}

If the susceptibility $\chi$ \textemdash \ and hence the coupling
$\varsigma$ \textemdash \ is independent of $x$, then the extended
system is invariant under spatial translations. Associated to this
symmetry is a local conservation law
\begin{equation}
  \partial_t p(x,t) + \partial_x T(x,t) \ = \ -f(x,t) \partial_x \phi(x,t) ,
\end{equation}
 for the wave momentum density
\begin{equation}
\begin{aligned}
  p(x,t) \ &= \ -\pi(x,t) \partial_x \phi(x,t) - \int_{\mathbb R}
  \theta(x,s,t) \partial_x \psi(x,s,t) \mathrm d s \\
  &= \ - \left \{ \partial_t \phi(x,t) + \Delta(x,t) \right \} \partial_x \phi(x,t)  -
  \int_{\mathbb R} \partial_t \psi(x,s,t) \partial_x \psi(x,s,t) \, \mathrm d s
  , \end{aligned}
\end{equation}
with wave momentum flux, called \emph{stress},
\begin{multline}
T(x,t) \ = \  E_0(x,t) + \Delta(x,t) \partial_t \phi(x,t)  \\ + \int
\left \{ (\partial_t \psi(x,s,t))^2 - (\partial_s \psi(x,s,t))^2
\right \} \mathrm d s .
\end{multline}
When the driving force vanishes, $f = 0$, the total wave momentum
\begin{equation}
P \ = \ \int_{\mathbb R} p(x,t) \, \mathrm d x
\end{equation}
is a conserved quantity.

The wave momentum density $p = p_0 + p_{hs}$ is again a sum of
contributions
\begin{align}\label{eq:p0}
p_0(x,t) \ &= \  - \left \{ \partial_t \phi(x,t) + \Delta(x,t) \right \} \partial_x \phi(x,t) \\
p_{hs}(x,t) \ &= \ -\int_{\mathbb R}
  \partial_t \psi(x,s,t) \partial_x \psi(x,s,t) \, \mathrm d s
  \label{eq:phs}
\end{align}
from the physical and hidden strings. Likewise we separate the
stress
\begin{equation}\label{eq:Ttotal}
T(x,t) \ = \ T_0(x,t) + T_{hs}(x,t)
\end{equation}
into two pieces,
\begin{align}\label{eq:T0}
T_0(x,t) \ &= \  E_0(x,t) + \Delta(x,t) \partial_t
\phi(x,t)  \\
T_{hs}(x,t) \ &= \ \frac{1}{2}\int \left \{ (\partial_t
\psi(x,s,t))^2 - (\partial_s \psi(x,s,t))^2 \right \} \mathrm d s
.\label{eq:Ths}
\end{align}
Then
\begin{align}
\partial_t p_0(x,t) + \partial_x T_0(x,t) \ &= \   -f(x,t) \partial_x \phi(x,t) + \partial_x
\Delta(x,t) \partial_t \phi(x,t)
  \\
\partial_t p_{hs}(x,t)  + \partial_x T_{hs}(x,t) \ &= \  - \partial_x \Delta(x,t)\partial_t \phi(x,t)  .
\end{align}

When we study eigenfunctions below, it will be convenient to work
with complex valued solutions. In the above expressions, terms which
are quadratic in the field variables should be modified in the
complex case by the replacement $a b \rightarrow \mathrm{Re} \,
\overline a b$.  That is
\begin{equation}
  \begin{aligned}
    E_0(x,t) \ &= \ \frac{1}{2} \left \{ |\partial_t \phi(x,t)|^2 + \gamma
    |\partial_x \phi(x,t)|^2 \right \} \\
    E_{hs}(x,t) \ &= \ \frac{1}{2}  \int_{\mathbb R} \left \{
    |\partial_t \psi(x,s,t)|^2 + |\partial_s \psi(x,s,t)|^2 \right \} \mathrm
    d s  \\
    J(x,t) \ &= \ - \gamma \mathrm{Re} \,\overline{\partial_t \phi(x,t)}
    \partial_x \phi(x,t) \\
p_0(x,t) \ &= \ -\mathrm{Re} \, \overline{\partial_t \phi(x,t)}
\partial_x \phi(x,t) - \mathrm{Re} \, \overline{\Delta(x,t)} \partial_x \phi(x,t) ,
  \end{aligned}
\end{equation}
etc.

\section{The eigenfunction equation}
It is useful and interesting to study steady state solutions to the
extended system \eqref{eq:extended}, for example solutions which are
periodic in time $\mathrm e^{-\mathrm i \omega t}
(\phi_\omega(x),\pi_\omega(x),$
$\psi_\omega(x,s),\theta_\omega(x,s))$. We refer to the spatial
component $\Phi_\omega(x,s) = (\phi_\omega(x), \pi_\omega(x),$
$\psi_\omega(x,s),\theta_\omega(x,s))$ of such a time periodic
solution as an eigenfunction for the linear system
\eqref{eq:extended} with eigenvalue $\omega$. Thus an eigenfunction
satisfies
\begin{equation}\label{eq:extendedeigenfunction}
\begin{aligned}
-\mathrm i \omega \phi_\omega(x)  \ &= \ f_\pi(x)  \\
-\mathrm i \omega \pi_\omega(x) \ &= \ \gamma \partial_x^2 \phi_\omega(x)\\
-\mathrm i \omega \psi_\omega(x,s) \ &= \ \theta_\omega(x,s) \\
-\mathrm i \omega \theta_\omega(x,s) \ &= \ \partial_s^2
\psi_\omega(x,s) + \varsigma(x,s) f_\pi(x) ,
\end{aligned}
\end{equation}
with
\begin{equation}
   f_\pi(x) + \int_{\mathbb R} \varsigma(x,s) \psi_\omega(x,s)\, \mathrm d s -
   \pi_\omega(x) \ = \ 0 .\label{eq:fpiTDDegenfunction}
\end{equation}

We see that the displacement of the hidden string at position $x$
satisfies
\begin{equation}\label{eq:hseigenvalue}
  - \partial_s^2 \psi_\omega(x,s)-\omega^2 \psi_\omega(x,s) \ = \  -\mathrm i
  \omega \varsigma(x,s) \phi_\omega(x) ,
\end{equation}
so
\begin{multline}\label{eq:hseigenvalue2}
\psi_\omega(x,s) \\
\begin{aligned}
   &= \ a(x) \cos(\omega s) + b(x) \sin(\omega s) -  \frac{\mathrm i \omega \phi_\omega(x)}{2
  \pi} \mathrm{P.V.}
  \int_{-\infty}^\infty \mathrm e^{-\mathrm i \sigma s}
  \frac{1}{\sigma^2 - \omega^2} \widehat \varsigma(x,\sigma) \, \mathrm d \sigma \\
    &= \ a(x) \cos(\omega s) + b(x) \sin(\omega s) + \frac{\mathrm i \phi_\omega(x)}{2}
  \int_{-\infty}^\infty \sin( \omega |s'-s|) \varsigma(x,s') \,
  \mathrm d s' ,
\end{aligned}
\end{multline}
with $a(x)$ and $b(x)$ undetermined functions of $x$. Here,
$\mathrm{P.V.}$ denotes the ``principle value'' integral,
\begin{multline}
  \mathrm{P.V.} \int_{-\infty}^\infty \mathrm e^{-\mathrm i \sigma s}
  \frac{1}{\sigma^2 - \omega^2} \widehat \varsigma(x,\sigma) \mathrm d
  \sigma \\
\begin{aligned}
 &= \ \lim_{\delta \downarrow 0} \int_{\{ \sigma \ :
\
  |\sigma^2 - \omega^2 | \ge \delta \}} \mathrm e^{-\mathrm i \sigma s}
  \frac{1}{\sigma^2 - \omega^2} \widehat \varsigma(x,\sigma) \, \mathrm d
  \sigma \\
  &= \ \lim_{\delta \downarrow  0} \frac{1}{2} \int_{\mathbb R} \mathrm e^{-\mathrm i \sigma s}
  \left [ \frac{1}{\sigma^2 - (\omega + \mathrm i \delta)^2}
  + \frac{1}{\sigma^2 - (\omega - \mathrm i \delta)^2} \right ]
  \widehat \varsigma(x,\sigma) \, \mathrm d
  \sigma .
\end{aligned}
\end{multline}

By \eqref{eq:hseigenvalue2}, we see that
\begin{multline}
  \int_{\mathbb R} \varsigma(x,s) \psi_\omega(x,s) \mathrm d s
  \ = \ a(x) \widehat \varsigma(x,\omega) \\ - \ \frac{\phi_\omega(x)}{2 \pi} \mathrm{P.V.}
  \int_{-\infty}^\infty
    \frac{\mathrm i \omega}{\sigma^2 - \omega^2} \widehat \varsigma(x,\sigma) \widehat \varsigma(x,-\sigma)\, \mathrm d \sigma .
\end{multline}
Recalling \eqref{eq:fundamentalFT}, that $\widehat
\varsigma(x,\sigma)\widehat \varsigma(x,-\sigma) = \widehat
D_\chi(x,\sigma) = 2\sigma \mathrm{Im} \, \widehat \chi(x,\sigma)$,
we see that the final term on the r.h.s.\ can be expressed
\begin{multline}
 -\frac{\phi_\omega(x)}{2 \pi} \mathrm{P.V.} \int_{-\infty}^\infty
    \frac{2 \mathrm i \omega }{\sigma^2 - \omega^2}
    \sigma \mathrm{Im} \, \widehat \chi(x,\sigma)  \, \mathrm d \sigma
  \\ \begin{aligned} &= \ - \frac{\mathrm i \phi_\omega(x)}{2 \pi} \mathrm{P.V.}
  \int_{-\infty}^\infty \left \{ \frac{1}{\sigma - \omega }
  + \frac{1}{-\sigma -\omega } \right \} \sigma \mathrm{Im} \, \widehat \chi(x,\sigma)\, \mathrm d
  \sigma \\
   &= \ -\mathrm i \omega \mathrm{Re}\, \widehat \chi(x,\omega) \phi_\omega(x),
  \end{aligned}
\end{multline}
where we have used the symmetry $\sigma \mathrm{Im}\, \widehat
\chi(x,\sigma) = - \sigma \mathrm{Im} \, \widehat \chi(x, -\sigma)$
and the ``Kramers-Kr\"onig relation''
\begin{equation}
  \omega \mathrm{Re} \, \widehat \chi(x,\omega)
   \ = \ \mathrm{P.V.} \frac{1}{\pi} \int_{-\infty}^\infty \frac{1}{\sigma -
   \omega} \sigma \mathrm {Im} \, \widehat \chi(x,\sigma) \,  \mathrm d
   \sigma .
\end{equation}

Plugging this result into \eqref{eq:fpiTDDegenfunction} and
combining the first two equations of
\eqref{eq:extendedeigenfunction} we find that the string
displacement $\phi$ solves the following elliptic problem
\begin{equation}\label{eq:eigenfunctionequation}
\gamma \partial_x^2 \phi_\omega(x) \ = \ -\omega^2(1 + \mathrm{Re}
\widehat \chi(x,\omega) ) \phi_\omega(x) - \mathrm i \omega a(x)
\widehat \varsigma(x,\omega).
\end{equation}
In analyzing \eqref{eq:eigenfunctionequation} we should distinguish
two cases:
\begin{enumerate}
\item $\widehat \varsigma(x,\omega) = 0$ for all $x$.
\item $\widehat \varsigma(x,\omega)$ non zero for $x$ in an open set.
\end{enumerate}
In the first case, which is non-generic, the medium described by the
hidden strings is not absorbing at the given frequency $\omega$,
that is $\widehat \chi(x,\omega)$ is real for every $x$.  Thus
\eqref{eq:eigenfunctionequation} is a very strong restriction on
$\phi_\omega$, namely that it should satisfy the Schr\"odinger
equation
\begin{equation}
(- \gamma \partial_x^2 +V(x)) \phi_\omega(x) = \lambda
\phi_\omega(x) ,
\end{equation}
with potential $V(x) =  - \omega^2\widehat \chi(x,\omega)$ and
spectral parameter $\lambda=\omega^2$. In the second case, the
physical string displacement may be decomposed as follows
\begin{equation}
  \phi_\omega(x) \ = \ \phi_\omega^{(1)}(x) + \phi_\omega^{(2)}(x) ,
\end{equation}
where $\phi_\omega^{(1)}(x)$ is any solution to the non-dissipative
eigenfunction equation
\begin{equation}
  \gamma \partial_x^2 \phi_\omega^{(1)}(x) \ = \ -\omega^2(1 + \mathrm{Re}
  \,
\widehat \chi(x,\omega) ) \phi_\omega^{(1)}(x) ,
\end{equation}
and $\phi_\omega^{(2)}$ is an \emph{arbitrary function with support
in $\{ x \ : \ \widehat \varsigma(x,\omega) \neq 0 \}$.} Indeed,
given $\phi_\omega^{(1)}$ and $\phi_\omega^{(2)}$ we need only
choose $a$ to be
\begin{equation}
  a(x)  \ = \ \mathrm i \frac{ \gamma \partial_x^2 \phi_\omega^{(2)}(x)
  + \omega^2 (1 + \mathrm{Re} \, \widehat \chi(x,\omega)
  )\phi_\omega^{(2)}(x)}{ \omega \widehat \varsigma(x,\omega)}
\end{equation}
to obtain a solution to \eqref{eq:eigenfunctionequation}.

\emph{Thus, the physical string displacement for an eigenmode $\phi
_{\omega }(x)$ may be chosen arbitrarily within that part of the
string which is absorbing at the given frequency }$\omega $. We want
to emphasize the significance of this fact, since the formal
eigenvalue problem
\begin{equation}
\gamma \partial _{x}^{2}\phi _{\omega }(x)\ =\ -\omega
^{2}(1+\widehat{\chi } (x,\omega ))\phi _{\omega }(x)
\end{equation}%
does not allow for such arbitrariness in the choice of $\phi
_{\omega }(x)$. The resolution to this apparent contradiction lies
in recognizing that a TDD system is an open part of a larger
conservative Hamiltonian system. It is only for the extended
Hamiltonian system that the eigenmodes $\Phi _{\omega }$ are
unambiguously defined with $e^{\mathrm{i}\omega t}\Phi _{\omega }$ a
stationary solution to the canonical Hamiltonian evolution equation.
The ``legitimate'' eigenmodes for the original TDD string consist of
projections $\phi_\omega$ of the eigenmodes $\Phi_\omega$ onto the
subspace of the physical string. Thus a TDD string, being an open
system has as many eigenmodes, i.e stationary solutions, as its the
minimal conservative extension, introduced and described in
\cite{FS1, FS2, FS3}. In particular, a finite-dimensional TDD system
typically has infinitely many stationary solutions and, hence,
infinitely many eigenmodes. Another view on the construction of
eigenmodes follows.

The eigenfunctions written down above, involving as they do
arbitrary excitations of the hidden strings, may not be relevant to
the dynamics \eqref{eq:extended} with the external force acting on
the physical string. Indeed, note that the effective equation
\eqref{eq:eigenfunctionequation} for $\phi_\omega$ does not depend on
the term $b(x) \sin(\omega s)$ appearing in \eqref{eq:hseigenvalue2}.  We shall see 
that eigenmodes with $b(x) \neq 0$ are not needed in the expansion of  a general solution to
\eqref{eq:extended} with a compactly supported external force. Essentially this is due to the fact that the configuration of the hidden strings remains symmetric throughout the evolution \eqref{eq:extended}. 

To proceed, it is
convenient to introduce the Fourier-Laplace transform
\begin{equation}
  \widetilde h(\zeta) \ = \ \int_{\mathbb R} \mathrm e^{\mathrm
  i \zeta t} h(t) \, \mathrm d t,
\end{equation}
defined for
\begin{enumerate}
\item $\zeta \in \mathbb C$ if $h \rightarrow 0$ super exponentially fast as  $|t| \rightarrow
\infty$, for instance if $h$ is compactly supported in time.
\item $\mathrm{Im} \zeta > 0$ if $h \rightarrow 0$ super exponentially fast as  $t \rightarrow
-\infty$.
\item $\mathrm{Im} \zeta < 0$ if $h \rightarrow 0$ super exponentially fast as  $t \rightarrow
\infty$.
\end{enumerate}
Taking the Fourier-Laplace transform of \eqref{eq:extended} results
in the following equations
\begin{equation}\label{eq:FLextended}
\begin{aligned}
- \mathrm i \zeta  \widetilde \phi(x,\zeta)  \ &= \ \widetilde f_\pi(x,\zeta)  \\
- \mathrm i \zeta  \widetilde\pi(x,\zeta) \ &= \ \gamma \partial_x^2 \widetilde \phi(x,\zeta) + \widetilde f(x,\zeta)\\
- \mathrm i \zeta  \widetilde \psi(x,s,\zeta) \ &= \ \widetilde \theta(x,s,\zeta) \\
- \mathrm i \zeta  \widetilde \theta(x,s,\zeta) \ &= \ \partial_s^2
\widetilde \psi(x,s,\zeta) + \varsigma(x,s) \widetilde
f_\pi(x,\zeta) ,
\end{aligned}
\end{equation}
with
\begin{equation}\label{eq:FLextendedfpi}
\widetilde f_\pi(x,\zeta) \ = \ \widetilde \pi(x,\zeta) -
\int_{-\infty}^\infty \varsigma(x,s) \widetilde \psi(x,s,\zeta) \,
\mathrm d s.
\end{equation}

Due to our consideration of solutions with the strings at rest at
$t=-\infty$, we expect the quantities in \eqref{eq:FLextended} to be
well defined only for $\mathrm{Im}\, \zeta
> 0$.  However, if the driving force is compactly supported in time
then $\widetilde f(x,\zeta)$ is defined for all $\zeta \in \mathbb
C$, and we may extend $\widetilde \phi(x,\zeta)$ to $\zeta \in
\mathbb C \setminus \mathbb R$ by solving \eqref{eq:FLextended}.
Following the eigenfunction analysis, we obtain, with $\varepsilon =
$ sign of $\mathrm{Im} \zeta$,
\begin{align}\label{eq:FLpsi}
  \widetilde \psi(x,s,\zeta) \ &=  \ - \frac{\mathrm i \zeta \widetilde \phi(x,\zeta)}{2
  \pi}
  \int_{-\infty}^\infty \mathrm e^{-\mathrm i \sigma s}
  \frac{1}{\sigma^2 - \zeta^2} \widehat \varsigma(x,\sigma) \, \mathrm d \sigma \\
  \notag &= \ \varepsilon  \frac{\widetilde \phi(x,\zeta)}{2}
  \int_{-\infty}^\infty \mathrm e^{\mathrm i \varepsilon \zeta |s'-s|} \varsigma(s',
  x) \,
  \mathrm d s', \\ \label{eq:FLphi}
  \gamma \partial_x^2 \widetilde \phi(x,\zeta) \ &= \ -\zeta^2(1 + \widehat
\chi(x,\varepsilon \zeta) ) \widetilde \phi(x,\zeta) + \widetilde
f(x,\zeta),
\end{align}
where $\widehat \chi(x, \zeta)$ is defined for $\mathrm{Im} \, \zeta
> 0$ as
\begin{equation}
\widehat \chi(x, \zeta) \ = \
  \int_0^\infty \mathrm e^{\mathrm i \tau \zeta} \chi(x, \tau) \,
  \mathrm d \tau .
\end{equation}

For a function $h$ vanishing at $+\infty$ or $-\infty$ we have the
Fourier inversion formula
\begin{equation}
  h(t) \ = \ \frac{1}{2 \pi } \int_{\mathbb R} \mathrm e^{-\mathrm
  i \omega t \mp \mathrm \eta t} \widetilde h(\omega \mp \mathrm i
  \eta) \, \mathrm d \omega , \quad \Bigl (\lim_{t = \pm \infty} h(t) = 0
  \Bigr ),
\end{equation}
with $\eta >0$ arbitrary.  Inverting the solution to
\eqref{eq:FLextended} with $\eta >  0$ or $< 0$ produces distinct
solutions to \eqref{eq:extended}: for $\eta > 0$ we obtain the
desired \emph{causal} solution with the strings at rest at $t=
-\infty$, while for $\eta < 0$ we obtain the \emph{anti-causal}
solution with the strings at rest at $t =+ \infty$.

In a certain sense we are only interested in the causal solution to
\eqref{eq:extended} and thus to the solution to
\eqref{eq:FLextended} only for $\zeta$ in the upper half plane.
However, since \eqref{eq:FLextended} involves a source term, this
solution, often called the \emph{resolvent}, is not directly
expressed as a superposition of eigenfunctions.  However, there is a
general procedure for decomposing the solution to
\eqref{eq:FLextended} into a superposition of eigenfunctions. Namely
for $\omega \in \mathbb R$ we define
\begin{multline}\label{eq:eigenfunctionasjump}
\left (\phi_\omega(x), \pi_\omega(x), \psi_\omega(x,s),
\theta_\omega(x,s) \right ) \\ = \ \lim_{\delta \rightarrow 0} \left
(\widetilde \phi(x,\omega + \mathrm i \delta) - \widetilde
\phi(x,\omega - \mathrm i \delta), \widetilde \pi(x,\omega + \mathrm
i \delta) - \widetilde \pi(x,\omega - \mathrm i \delta), \right . \\
\left . \widetilde \psi(x,s,\omega + \mathrm i \delta) - \widetilde
\psi(x,s,\omega - \mathrm i \delta), \widetilde \theta(x,s,\omega +
\mathrm i \delta) - \widetilde \phi(x,s,\omega - \mathrm i \delta)
\right ).
\end{multline}
Since $\widetilde f(x,\zeta)$ is continuous at each $\zeta = \omega
\in \mathbb R$, it follows from \eqref{eq:FLextended} that
$(\phi_\omega,\pi_\omega,$ $\psi_\omega,\theta_\omega)$ is an
eigenfunction for each $\omega$. To recover the resolvent for
$\zeta$ in the upper half plane from the eigenfunctions
\eqref{eq:eigenfunctionasjump}, suppose that the external force is
supported in the set $\{ t \ : \ t > t_0 \}$.  Then
\begin{equation}
  |\widetilde f(x,\zeta) | \ \le \ C \ \mathrm e^{-\mathrm{Im} \,
  \zeta t_0}, \quad (\mathrm{Im} \, \zeta > 0) .
\end{equation}
Thus $\exp (- \mathrm i \zeta t_0) f(x,\zeta)$ is bounded in the
upper half plane, and by analyticity we have,\footnote{For a general
linear system the limit on the right hand side of
\eqref{eq:eigenfunctionasjump} is a defined only as a vector valued
measure and the integral on the right hand side of
\eqref{eq:superposition} should be interpreted as the integral of
$1/(\omega - \zeta)$ against this measure.  For the systems of
coupled strings considered here, however, these boundary measures
are always absolutely continuous, so \eqref{eq:superposition} holds
with $(\phi_\omega,\pi_\omega,\psi_\omega,\theta_\omega)$ defined
for almost every $\omega$ by \eqref{eq:eigenfunctionasjump}.}
\begin{multline}\label{eq:superposition}
\left  (\widetilde \phi(x,\zeta), \widetilde \pi(x,\zeta),
\widetilde
  \psi(x,s,\zeta), \widetilde \theta(x,s,\zeta) \right ) \\ = \
  \frac{1}{2 \pi \mathrm i} \int_{\mathbb R} \frac{\mathrm e^{\mathrm i (\zeta - \omega) t_0} }{\omega - \zeta}
  \left (\phi_\omega(x), \pi_\omega(x), \psi_\omega(x,s),
\theta_\omega(x,s) \right )  \, \mathrm d \omega, \quad
(\mathrm{Im}\, \zeta > 0 )
\end{multline}
expressing the solution to \eqref{eq:FLextended} as a superposition
of eigenfunctions. (There is a similar formula for the advanced
solution with $\mathrm{Im} \, \zeta < 0$, involving an upper bound
$t_1$ on the support of the external force.)

Now let us fix $\omega$ and suppose that we have an eigenfunction of
the form \eqref{eq:eigenfunctionasjump}. Then by \eqref{eq:FLpsi}
and \eqref{eq:FLphi},
\begin{align}
  \psi_\omega(x,s) \ &= \ \frac{\mathrm i \phi_\omega(x)}{2}
  \int_{-\infty}^\infty \sin( \omega |s'-s|) \varsigma(x,s') \,
  \mathrm d s' +  \frac{g_\omega(x)}{2} \cos(\omega s) \widehat \varsigma (x,\omega) \\
  \gamma \partial_x^2 \phi_\omega(x) \ &= \ -\omega^2(1 + \mathrm{Re} \, \widehat
\chi(x,\omega) ) \phi_\omega(x) -  \mathrm i \omega^2 \mathrm{Im} \,
\widehat \chi(x,\omega) g_\omega(x) ,
\label{eq:spectraleigenfunction}
\end{align}
with
\begin{equation}
  g_\omega(x) \ = \ \lim_{\delta \rightarrow 0} \left \{
    \widetilde \phi(x,\omega + \mathrm i \delta) + \widetilde \phi(x,\omega - \mathrm i
    \delta) \right \} .
\end{equation}
In other words, the eigenfunctions which appear in the expansion
\eqref{eq:superposition} are of the form
\eqref{eq:eigenfunctionequation} with $b(x)=0$.  We refer to such
eigenfunctions as \emph{spectral} eigenfunctions. By
\eqref{eq:superposition}, the spectral eigenfunctions form a
complete set for the description of the dynamics of the extended
system \eqref{eq:extended}. Thus, the freedom to choose $g_\omega$
in \eqref{eq:spectraleigenfunction} provides us with a rich enough
family of solutions to describe the dynamics of the TDD string.

Note that for $\zeta = \omega \in \mathbb R$, a  solution to  \eqref{eq:FLextended} satisfies the
eigenmode equation away from the spatial support of the driving
force.  Thus, one may try to produce  eigenfunctions via a solution to \eqref{eq:FLextended}
with a ``source at infinity,'' that is as a limit
of a sequence of solutions to \eqref{eq:FLextended} with driving forces supported farther and farther from the origin. Depending on the susceptibility and the sequence of driving forces, the limit may or may exist and be a non zero.
However, when this procedure leads to a non-trivial limit, the resulting eigenmodes are of the form \eqref{eq:spectraleigenfunction} with a
special choice of the arbitrary function $g_\omega$. If we solve \eqref{eq:FLextended} in the upper or lower half planes, we obtain in this way
\begin{enumerate}
  \item The \emph{causal eigenfunctions} with $g_\omega(x) = \phi_\omega(x)$,
  \begin{align}\label{eq:causalhs}
  \psi_\omega(x,s) \ &= \ \frac{\phi_\omega(x)}{2}
  \int_{-\infty}^\infty \mathrm e^{\mathrm i \omega |s'-s|} \varsigma(s',
  x) \,
  \mathrm d s' \\
  \gamma \partial_x^2 \phi_\omega(x) \ &= \ -\omega^2(1 + \widehat
    \chi(x,\omega) ) \phi_\omega(x), \label{eq:causaleigenfunction}
\end{align}
  corresponding to a driving force in the distant past and the
  causal boundary condition with the strings at rest at $t=-\infty$.
  \item The \emph{anti-causal eigenfunctions} with $g_\omega(x) = -
  \phi_\omega(x)$,
  \begin{align}
  \psi_\omega(x,s) \ &= \ \frac{\phi_\omega(x)}{2}
  \int_{-\infty}^\infty \mathrm e^{-\mathrm i \omega |s'-s|} \varsigma(s',
  x) \,
  \mathrm d s' \\
  \gamma \partial_x^2 \phi_\omega(x) \ &= \ -\omega^2(1 + \overline{\widehat
    \chi(x,\omega)} ) \phi_\omega(x), \label{eq:anticausaleigenfunction}
    \end{align}
    corresponding to a driving force in the
  distant future and the
  anti-causal, or advanced, boundary condition with the strings at
  rest at $t = +\infty$.
\end{enumerate}

Note that the causal eigenfunction equation \eqref{eq:causaleigenfunction} is simply the \emph{formal} eigenvalue problem obtained from the time Fourier transform  of the TDD string equation \eqref{eq:TDDwaveeqn}. Similarly, the anti-causal eigenfunction equation \eqref{eq:anticausaleigenfunction} is the formal eigenvalue problem obtained from the time reversal of \eqref{eq:TDDwaveeqn}.   However, the solutions to \eqref{eq:causaleigenfunction} and \eqref{eq:anticausaleigenfunction} are very special in nature and need not provide a complete set for the expansion of solutions to \eqref{eq:TDDwaveeqn}.  Indeed, it can happen that there are \emph{no} physically relevant causal or anti-causal eigenfunctions.   For instance, in a homogeneous string the only solutions to \eqref{eq:causaleigenfunction} and \eqref{eq:anticausaleigenfunction} are  exponentially growing as $x \rightarrow + \infty$ or $x \rightarrow - \infty$, and are thus irrelevant.\footnote{One can use the abstract theory of Gelfand triples and generalized eigenfunction expansions to show that only those eigenfunctions which grow at infinity slower than $(1 + |x|)^{1 + \delta}$ with arbitrary $\delta$ are relevant in the eigenvalue expansion of \eqref{eq:extended}.}  Nonetheless, there are of course plenty of spectral eigenfunctions coming from the extended system. In section \ref{sec:planewave} below we construct a rich family of bounded plane wave solutions for a homogeneous string, using \eqref{eq:spectraleigenfunction} with suitable choices for the source term $g_\omega$.

This is not to say that the causal and anti-causal eigenfunctions are always irrelevant.  That depends on the physics, and indeed the formal eigenvalue problem can be useful in the right context. For instance, if dissipation is restricted to a proper subset of the string, or more generally if $\mathrm{Im} \, \widehat \chi(x,\omega)$ falls off sufficiently fast as $x \rightarrow \pm \infty$, then among the causal eigenfunctions are scattering solutions which describe the reflection and transmission of a plane wave emitted from a source at $x = \pm \infty$.  In the complete family of spectral eigenfunctions these solutions are very special, however they are precisely those solutions needed to analyze the extended dynamics \eqref{eq:extended} for a driving force situated very far from the dissipative portion of the string. In section \ref{sec:scattering} we illustrate the role of  causal eigenfunctions in scattering theory by computing the scattering modes for a string in which dissipation is restricted to $x > 0$.

\section{Energy flux in an eigenfunction}
The energy density $\mathcal E$ in an eigenfunction, at a point $x$
with $\widehat \varsigma(x,\omega) \neq 0$, is typically infinite
due to the contribution from the hidden strings. This result is to
be expected physically as the eigenfunction represents the idealized
steady propagation of a monochromatic wave through an absorbing
medium with infinite heat capacity. However, due to the decoupling
between the hidden strings, energy can flow only through the
physical string and we expect the energy flux $J$ to be finite.
Furthermore as we shall see $\partial_x J$, which by
\eqref{eq:localconservation} is formally equal to $-
\partial_t \mathcal E$, can be non-zero at a point $x$, in which
case the eigenfunction is a steady state in which energy is
dissipated to or absorbed from the medium at $x$ at a constant rate.

By \eqref{eq:energyflux}, the energy flux for an eigenfunction is
\begin{equation}\label{eq:energyfluxef}
  J(x,t) \ = \ - \gamma \mathrm{Re} \left \{  \mathrm i \omega
  \overline{\phi_\omega(x)} \partial_x \phi_\omega(x) \right \} \ = \ \gamma
  \omega
  \mathrm{Im} \, \overline{\phi_\omega(x)} \partial_x \phi_\omega(x) .
\end{equation}
Thus, by \eqref{eq:spectraleigenfunction},
\begin{equation}
\begin{aligned}
  - \partial_x J(x) \  &= \ - \omega
  \mathrm{Im} \,
  \overline{\phi_\omega(x)} \gamma \partial_x^2 \phi_\omega(x) \\
  &= \ \omega^3 \mathrm{Im} \, \widehat \chi(x,\omega) \mathrm{Re}\left \{
   \overline{\phi_\omega(x)}g_\omega(x) \right \}
  .
\end{aligned}
\end{equation}
for a spectral eigenfunction, where $g_\omega(x)$ is the arbitrary
function describing the excitation of the medium, as represented by
the hidden strings. The energy density of the physical string
\begin{equation}\label{eq:E0eigenfunction}
  E_0(x) \ = \ \frac{1}{2} \left \{ \omega^2 |\phi_\omega(x)|^2 + \gamma |
  \partial_x \phi_\omega(x)|^2 \right \} ,
\end{equation}
is constant in time, and there is a constant rate of dissipation at
each $x$ with $-\partial_x J(x) \neq 0$,
\begin{equation}
  \partial_t \mathcal E(x) \ = \ \partial_t E_{hs}(x) \ = \
  \omega^3 \mathrm{Im} \, \widehat \chi(x,\omega)  \mathrm{Re} \, \left \{
   \overline{\phi_\omega(x)}g_\omega(x) \right \} .
\end{equation}
Thus the eigenfunction represents a steady state situation in which
energy is flowing into or out of the hidden strings at a constant
rate at each point $x$ with $\mathrm{Im} \, \widehat \chi(x,\omega)
\neq 0$.

For a general spectral eigenfunction
\eqref{eq:spectraleigenfunction}, the dissipation $\partial_t
\mathcal E(x)$ may be positive or negative, however for a causal
eigenfunction \eqref{eq:causaleigenfunction}, with $g_\omega =
\phi_\omega$, we have
\begin{equation}
\partial_t \mathcal E(x) \ = \ \partial_t E_{hs}(x) \ = \
  \omega^3 \mathrm{Im} \,\widehat \chi(x,\omega)  |\phi_\omega(x)|^2 \ > \ 0 , \quad \text{(causal eigenfunction)}
\end{equation}
corresponding to a steady dissipation of energy from the physical
string into the medium, represented by the hidden strings.
Similarly, for an anti-causal eigenfunction
\eqref{eq:anticausaleigenfunction} there is a steady flux of energy
\emph{out of the medium and into the physical string}
\begin{equation}
\partial_t E_{hs}(x) \ = \ - \omega^3 \mathrm{Im} \, \widehat
\chi(x,\omega) |\phi_\omega(x)|^2 \ < \ 0, \quad \text{(anti-causal
eigenfunction).}
\end{equation}

\section{Plane waves and momentum flux in an homogeneous system}\label{sec:planewave}
Suppose that the susceptibility $\chi(x,\tau) = \chi(\tau)$ is
constant for the whole range of $x \in \mathbb R$.  Then it is
interesting to look for plane wave eigenfunctions $\mathrm
e^{\mathrm{i} (kx - \omega t)} (\phi_0,\pi_0,$
$\psi_0(s),\theta_0(s))$ with $\phi_0,\pi_0$ constants and
$\psi_0(s), \theta_0(s)$ independent of $x$.

A first observation is that there are no causal or anti-causal
eigenfunctions of this form, at least at frequencies with
$\mathrm{Im}\, \widehat \chi(\omega) > 0$. Indeed a causal solution
would satisfy
\begin{equation}\label{eq:causalplanewave} - \gamma
k^2 \phi_0\ = \ -\omega^2(1 + \widehat \chi(\omega) ) \phi_0 ,
\end{equation}
and the only non-trivial solutions to this equation are
exponentially growing as $x \rightarrow \pm \infty$ unless $\widehat
\chi(\omega)$ is real. Such evanescent waves play a role in
constructing scattering states for inhomogeneous systems but are
physically irrelevant in the homogeneous system.

Thus to find plane wave solutions it is necessary to look beyond
causal eigenfunctions.  For a plane wave spectral eigenfunction, see
\eqref{eq:spectraleigenfunction}, we have
\begin{align}
- \gamma k^2 \phi_0 \ &= \ - \mathrm i \omega^2 g_0 \mathrm{Im} \,
\widehat
\chi(\omega)  - \omega^2(1 + \mathrm{Re} \widehat \chi(\omega) ) \phi_0 , \\
  \psi_0(s) \ &= \  \frac{g_0}{2} \cos(\omega s) \widehat \varsigma(\omega) + \frac{\mathrm i \phi_0}{2}
  \int_{-\infty}^\infty \sin( \omega |s'-s|) \varsigma(s') \,
  \mathrm d s' ,
\end{align}
with $g_0$ an arbitrary constant describing the excitation of the
hidden strings. A bounded solution results only for $k$ real, so
\begin{equation}
g_0 = \mathrm i \alpha \phi_0, \quad \alpha \in \mathbb R,
\end{equation}
must be a pure imaginary multiple of $\phi_0$. Furthermore, we must
have
\begin{equation}
  1 + \mathrm{Re} \widehat \chi(\omega)-  \alpha \mathrm{Im} \widehat \chi(\omega)
  \ \ge \ 0.
\end{equation}
Then setting
\begin{equation}
 \label{eq:dispersion} \gamma k^2  \ = \  \omega^2 \left (
 1 + \mathrm{Re} \widehat \chi(\omega)-  \alpha \mathrm{Im} \widehat \chi(\omega) \right ) ,
\end{equation}
we obtain a non-trivial plane wave solution.

If the medium is not dissipative at frequency $\omega$, so
$\mathrm{Im} \, \widehat \chi(\omega)=0$, there is a
\emph{dispersion} relation between $k$ and $\omega$
\begin{equation}\label{eq:dispersionrelation}
\gamma k^2 \ = \ \omega^2(  1 + \mathrm{Re} \, \widehat \chi(\omega)
), \quad (\mathrm{Im} \widehat \chi(\omega) = 0 ).
\end{equation}
At frequencies $\omega$ with non-trivial dissipation, that is
$\mathrm{Im} \widehat \chi(\omega) \neq 0$, there is \emph{no
relation between $k$ and $\omega$.}  Indeed $k$ may be chosen
arbitrarily provided we take
\begin{equation}
\alpha = \frac{1 + \mathrm{Re} \, \widehat \chi(\omega)}{\mathrm{Im}
\, \widehat \chi(\omega)} - \frac{\gamma k^2 }{\omega^2 \mathrm{Im}
\,  \widehat \chi(\omega)}.
\end{equation}

Observe that \eqref{eq:dispersionrelation} is essentially the Fourier transform in time and space of the formal eigenvalue problem \eqref{eq:causalplanewave} in the case that $\mathrm{Im} \, \widehat \chi(\omega) = 0$.   The lack of a dispersion relation between $k$ and $\omega$ in a dissipative medium highlights the failure of the formal eigenvalue problem in this context.  Indeed, we have seen that the formal eigenvalue problem \eqref{eq:causalplanewave} has \emph{no physically relevant solutions} in a homogeneous medium.  Nonetheless,  there is a complete eigenvalue basis for the extended system \eqref{eq:extended} consisting of the plane wave solutions derived above.

By \eqref{eq:energyfluxef}, the energy flux for a plane wave
\begin{equation}
J \ = \ \gamma \omega k |\phi_0|^2
\end{equation}
is constant.  Thus $\partial_x J= 0$ and plane wave solutions
represent a steady state flow without dissipation of energy.  More
precisely, the dissipation of energy to the medium, as described by
the hidden strings, is exactly balanced by the energy emitted from
the medium. The energy density of the physical string is
\begin{equation}\label{eq:PWE0}
E_0 \ = \ \left ( \omega^2 + \gamma k^2 \right ) |\phi_0|^2
\end{equation}
and the energy density of the medium, described by the hidden strings, is of course infinite.

As the system is homogeneous, we can also consider the wave momentum
density  and stress in a plane wave eigenfunction. By \eqref{eq:p0}
the wave momentum density of the physical string is
\begin{equation}
  p_0 \ = \ \omega k |\phi_0|^2 - k \mathrm{Im} \,\overline{\Delta_0} \phi_0 ,
\end{equation}
where
\begin{equation}
  \Delta_0 \ = \ \int_{\mathbb R} \varsigma(s) \psi_0(s)\,  \mathrm d s \ = \ \mathrm i \omega \phi_0 \left
  ( \alpha
  \mathrm{Im} \, \widehat \chi(\omega) - \mathrm{Re} \, \widehat
  \chi(\omega) \right ).
\end{equation}
Thus
\begin{equation}
  p_0 \ = \ \omega k \left \{ 1 + \mathrm{Re} \, \widehat \chi(\omega) - \alpha \mathrm{Im} \, \widehat \chi(\omega)
  \right \} |\phi|^2 \ = \ \frac{\gamma k^3}{\omega} |\phi_0|^2 .
\end{equation}
The wave momentum density of the hidden strings is infinite, since
\begin{equation}
  p_{hs} \ = \ \omega k \int_{\mathbb R} |\psi_0(s)|^2 \, \mathrm d s
\end{equation}
by \eqref{eq:phs}.

Of more interest is the stress, which by (\ref{eq:Ttotal},
\ref{eq:T0}) is
\begin{equation}
  T \ = \ E_0  + \omega \mathrm{Im} \overline \Delta_0 \phi_0 + T_{hs},
\end{equation}
where $E_0$ given by \eqref{eq:PWE0} and $T_{hs}$ is formally
\begin{equation}\label{eq:formalThs}
T_{hs} \ = \  \frac{1}{2} \int_{-\infty}^\infty \left \{ \omega^2
|\psi_0(s)|^2 - |\partial_s \psi_0(s)|^2 \right \} \mathrm d s.
\end{equation}
The integral on the r.h.s.\ of \eqref{eq:formalThs} does not
converge absolutely, but we can regularize it by defining $T_{hs} =
\lim_{\delta \rightarrow 0} T_{hs}(\delta)$ with
\begin{equation}\label{eq:Thsepsilon}
T_{hs}(\delta) \ = \  \frac{1}{2} \int_{-\infty}^\infty \mathrm
e^{-\delta |s|} \left \{ \omega^2 |\psi_0(s)|^2 - |\partial_s
\psi_0(s)|^2 \right \} \mathrm d s.
\end{equation}
To evaluate the integral in \eqref{eq:Thsepsilon} it is useful to
write
\begin{equation}
\begin{aligned}
  |\partial_s
\psi_0(s)|^2 \ &= \ \frac{1}{2} \partial_s^2 |\psi_0(s)|^2 -
\mathrm{Re}\,
\overline{\psi_0(s)} \partial_s^2 \psi_0(s) \\
&= \ \frac{1}{2} \partial_s^2 |\psi_0(s)|^2 + \omega^2 |\psi_0(s)|^2
+ \omega \varsigma(s) \mathrm{Im}\, \overline{\psi_0(s)} \phi_0 ,
\end{aligned}
\end{equation}
where we have used \eqref{eq:hseigenvalue}. Thus
\begin{equation}
\begin{aligned}
  T_{hs}(\delta) \ &= \ - \frac{1}{2} \int_{-\infty}^\infty \mathrm
e^{-\delta |s|} \left \{ \frac{1}{2} \partial_s^2 |\psi_0(s)|^2 +
\omega \varsigma(s) \mathrm{Im} \, \overline{\psi_0(s)} \phi_0 \right \}  \, \mathrm d s\\
&= \  \frac{\delta}{2} |\psi_0(0)|^2 - \frac{\delta^2}{4}
\int_{-\infty}^\infty  \mathrm e^{-\delta |s|} |\psi_0(s)|^2 \,
\mathrm d s - \frac{1}{2}\omega \int_{-\infty}^\infty  \mathrm
e^{-\delta |s|} \varsigma(s) \mathrm{Im} \, \overline{\psi_0(s)}
\phi_0 \, \mathrm d s.
\end{aligned}
\end{equation}
It follows that
\begin{equation}
T_{hs} \ = \
  \lim_{\delta \rightarrow 0} T_{hs}(\delta) \ = \ - \frac{1}{2} \omega \int_{-\infty}^\infty
  \varsigma(s) \mathrm{Im} \, \overline{\psi_0(s)} \phi_0 \, \mathrm d s \ = \
  - \frac{1}{2} \omega \mathrm{Im} \, \overline{\Delta_0} \phi_0 ,
\end{equation}
if, for instance, $|\varsigma(s)|$ is integrable. Thus
\begin{equation}
\begin{aligned}
T \ &= \
E_0 + \frac{1}{2} \omega \mathrm{Im} \, \overline{\Delta_0} \phi_0 \\
&= \ \frac{1}{2} \left ( \omega^2(1+ \mathrm{Re} \, \widehat
\chi(\omega) -  \alpha \mathrm{Im} \, \widehat \chi(\omega) ) +
\gamma k^2 \right ) |\phi_0|^2 \ = \ \gamma k^2 |\phi_0|^2 ,
\end{aligned}
\end{equation}
which is identical to the result for the undamped string!

\section{Scattering eigenfunctions for a semi-infinite medium}\label{sec:scattering} To
close, we would like to illustrate the power of the above method by
unambiguously constructing the scattering states representing an
idealized description of the following experiment.  Imagine that we
have an long, i.e., infinite, string whose right half ($x >0$) is
subject to dispersion and dissipation with susceptibility
$\chi(\tau)$. That is, the the susceptibility for the whole string
is
\begin{equation}
  \chi(x,\tau) \ = \ \begin{cases} 0 & x < 0 \\
  \chi(\tau) & x > 0 \end{cases}.
\end{equation}
Suppose we drive the left end of the string, at $x=-\infty$,
periodically with frequency $\omega$, sending an incoming wave
toward the TDD right half of the string.  After some time a steady
state is reached in which a certain fraction of the incoming wave is
absorbed by the dissipative half of the string and a certain
fraction is reflected.

The steady state eigenfunction describing the above experiment is a
causal eigenfunction since the source is at $x=-\infty$ and in the
distant past. Thus by \eqref{eq:causaleigenfunction}, the string
displacement satisfies
\begin{equation}\label{eq:causalhalfline}
  \gamma \partial_x^2 \phi_\omega(x) \ = \ - \omega^2 \Bigl (1 + I[x>0]
  \widehat \chi(\omega) \Bigr ) \phi_\omega(x) ,
\end{equation}
where $I[x>0]=1$ for $x > 0$ and $0$ otherwise. Indeed, this is the
naive equation that one might right down for the scattering states
of a TDD string.  However, it is only within the context of the
extended Hamiltonian system that we understand this to be only one
of many possible eigenfunction equations, the choice of which is
dictated by the physics under consideration, namely a source at
spatial infinity in the distant past.

By \eqref{eq:causalhalfline}, we have
\begin{equation}\label{eq:leftef}
  \phi_\omega(x) \ = \  \mathrm e^{\mathrm i k_< x} + r \mathrm
  e^{-\mathrm i k_< x}  , \quad k_< = \frac{\omega}{\sqrt{\gamma}}, \quad (x < 0)
\end{equation}
on the left half of the string, up to an over all multiple which we
fix to be $1$ without loss. In \eqref{eq:leftef} the term $\mathrm
e^{\mathrm i k_< x}$ is the incoming wave and $r \mathrm e^{-\mathrm
i k_> x}$ is the reflected wave.

To find the reflection coefficient $r$ we need to solve for the form
of the eigenfunction on the right half of the string. Again by
\eqref{eq:causalhalfline}, we have
\begin{equation}
  \gamma \partial_x^2 \phi_\omega(x) \ = \ - \omega^2 ( 1 + \widehat
  \chi(\omega) ) \phi_\omega(x), \quad (x > 0)
\end{equation}
for $x > 0$. Furthermore, since we require the solution to be
bounded this determines $\phi$ uniquely up to an over all multiple
\begin{equation}
  \phi_\omega(x) \ = \ v \mathrm e^{\mathrm i k_> x } ,
\end{equation}
where $v$ is an, as yet, undetermined transmission coefficient and
\begin{equation}
  k_> \ = \ \sqrt[+]{\frac{\omega^2}{\gamma} ( 1 + \widehat
  \chi(\omega) )} .
\end{equation}
Here $\sqrt[+]{z}$ denotes the unique square root of $z$ in the
upper half plane.\footnote{If $\widehat \chi(\omega)$ is real and
$(1 + \widehat \chi(\omega))
> 0 $ then the medium transmits at frequency $\omega$ and $k_>$
should be determined as the limit
$$
k_> \ = \ \lim_{\delta \downarrow 0} \sqrt[+]{\frac{(\omega+ \mathrm
i \delta)^2}{\gamma} ( 1 + \widehat
  \chi(\omega + \mathrm i \delta) )}.$$ } Since $\mathrm{Im} \, k_>
  > 0$, at least if $\mathrm{Im} \, \widehat \chi(\omega) \neq 0$,
  we see that $\phi_\omega(x)$ decays exponentially in the
  dissipative half of the physical string ($x > 0$).
  The excitation of the hidden strings, which are restricted to $x >
  0$, is given by \eqref{eq:causalhs}:
\begin{equation}
  \psi_\omega(x,s) \ = \ \frac{v}{2} \mathrm e^{\mathrm i k_> x }
  \int_{-\infty}^\infty \mathrm e^{\mathrm i \omega |s-s'|}
  \varsigma(s') \, \mathrm d s', \quad (x > 0).
\end{equation}

To determine $v$ and $r$ we note that the eigenfunction equation
\eqref{eq:causalhalfline} forces $\phi_\omega$ and $\partial_x
\phi_\omega$ to be continuous functions of $x$, in particular at $x
=0$. Thus,
\begin{align}
  \lim_{x \uparrow 0} \phi_\omega(x) \ = \ 1+ r \ &= \ v  \ = \ \lim_{x \downarrow 0} \phi_\omega(x) \\
  \lim_{x \uparrow 0} \partial_x \phi_\omega(x) \ = \ k_< ( 1- r) \ &= \ k_>
  v \ = \
  \lim_{x \downarrow 0} \partial_x \phi_\omega(x).
\end{align}
We conclude that
\begin{equation}\label{eq:transrefl}
  v \ = \ \frac{2}{1 + \rho(\omega) } , \quad r \ = \ \frac{1 -
  \rho(\omega)}{1 + \rho(\omega) } ,
\end{equation}
where
\begin{equation}
  \rho(\omega) \ = \ \frac{k_>}{k_<} \ = \ \begin{cases}
    \sqrt[+]{1 + \widehat \chi(\omega)} & \omega > 0, \\ -
    \sqrt[+]{1 + \widehat \chi(\omega)} & \omega < 0.
  \end{cases}
\end{equation}

It is useful to consider some general properties of $\rho(\omega)$
and the reflection and transmission coefficients. To begin note that
since $\omega \mathrm{Im} \,\widehat \chi(\omega) \ge 0$ we have
\begin{equation}
  \omega \mathrm{Re} \sqrt[+]{1 + \widehat \chi(\omega)} \ \ge \ 0 ,
\end{equation}
and thus
\begin{equation}
  \mathrm{Re} \, \rho(\omega) \ \ge \ 0 .
\end{equation}
This implies that
\begin{equation}\label{eq:1minmodr2}
  1 - |r|^2 \ = \ \frac{4}{1 + |\rho(\omega)|^2} \mathrm{Re} \,
  \rho(\omega) \ = \ |v|^2 \mathrm{Re} \, \rho(\omega) \ = \ |v|^2 \frac{\mathrm{Re} k_ >}{k_<} \ \ge \ 0 ,
\end{equation}
so, in particular,
\begin{equation}
  |r| \ \le \ 1.
\end{equation}

Eq.\ \eqref{eq:1minmodr2} expresses the continuity of the energy
flux $J(x)$ at $x= 0$, since by \eqref{eq:energyfluxef}
\begin{equation}
\begin{aligned}
  J(x) \ &= \ \gamma \omega \mathrm{Im} \,  \overline{\phi_\omega(x)} \partial_x
  \phi_\omega(x) \\ &= \ \gamma \omega \begin{cases} k_< (1 - |r|^2)  & x < 0 \\
  \mathrm{Re} k_> |v|^2 \mathrm e^{- 2 \mathrm{Im} k_> x} & x > 0
  \end{cases}\\
  &= \ \sqrt{\gamma} \omega^2 (1 - |r|^2) \begin{cases} 1 & x < 0 , \\ \mathrm e^{-2 \mathrm{Im} k_> x} & x >
  0 .
  \end{cases}
\end{aligned}
\end{equation}
Furthermore, we see that the energy flux is non-negative,
representing a flow of energy from the source at $x = -\infty$, and
the rate of dissipation, $-\partial_x J(x) \neq 0$, is non-zero for
every $x > 0$,
\begin{equation}
   -
  \partial_x J(x) \ = \ 2
  \sqrt{\gamma} \omega^2 (1 - |r|^2) \mathrm{Im} \,
  k_> \mathrm e^{-2 \mathrm{Im} \, k_> x}, \quad (x > 0 ).
\end{equation}

Finally, we note that the energy density of the physical string is
\begin{equation}
\begin{aligned}
  E_0(x) \ &= \ \frac{1}{2} \left \{ \omega^2 |\phi_\omega(x)|^2 + \gamma
  |\partial_x \phi_\omega(x)|^2 \right \} \\ &= \ \begin{cases} \omega^2 (1 +
  |r|^2) & x < 0,  \\
  \omega^2  \frac{1}{2} \left ( 1 + |1 +  \widehat \chi(\omega)|
  \right ) |v|^2 \mathrm e^{-2 \mathrm{Im} \, k_> x} & x > 0,
  \end{cases} \\
  &= \ \omega^2 (1 + |r|^2) \begin{cases} 1 & x < 0 , \\ \mathrm e^{-2 \mathrm{Im} \, k_> x} & x >
  0 ,
  \end{cases}
\end{aligned}
\end{equation}
where we have noted that by \eqref{eq:transrefl}
\begin{equation}
  1 + |r|^2 \ = \ \frac{1 + |\rho(\omega)|^2}{2} |v|^2 \ = \ \frac{1
  + |1 +  \widehat \chi(\omega)|}{2} |v|^2 .
\end{equation}

In summary, the scattering eigenmode
\begin{align}
  \phi(x,t) \ &= \  \begin{cases}
    \mathrm e^{\mathrm i \omega \left (\frac{x}{\sqrt{\gamma}} -t \right) } + r \mathrm
  e^{-\mathrm i \omega  \left (\frac{x}{\sqrt{\gamma}} +t \right)} & x< 0 \\
  v \mathrm e^{\mathrm i \omega \left (  \frac{ \rho(\omega)x}{\sqrt{\gamma}}-t \right )} & x > 0 ,
  \end{cases} \\
  \psi(x,s,t) \ &= \ \frac{v}{2}
  \mathrm e^{\mathrm i \omega \left (  \frac{ \rho(\omega)x}{\sqrt{\gamma}}-t \right ) }
  \int_{\mathbb R} \mathrm e^{\mathrm i \omega |s- s'|}
  \varsigma(s') \, \mathrm d s', \quad (x > 0),
\end{align}
describes a steady state in which the incoming wave is partially
reflected, with the remainder an evanescent transmitted wave that
penetrates the dissipative part of the string with an exponential
profile resulting in an excitation of the hidden strings accounting
for dispersion and dissipation. The total rate of dissipation in the
TDD portion of the string is
\begin{equation}
- \int_0^\infty \partial_x J(x)
  \mathrm d x
  \ = \ J(0) \ = \  \sqrt{\gamma} \omega^2(1 - |r|^2).
\end{equation}

\subsection*{Acknowledgment} The effort of A.\ Figotin was sponsored
by the Air Force Office of Scientific Research, Air Force Materials
Command, USAF, under grant number FA9550-04-1-0359. J.\ Schenker
received travel and materials support under the aforementioned USAF
grant and was supported by a National Science Foundation
post-doctoral fellowship.

\end{document}